\documentclass[aps,groupedaddress, showkeys, showpacs]{revtex4}

\usepackage[small]{caption}
\usepackage{amsmath,amsfonts,amscd,amssymb,epsf, epsfig}\usepackage{graphicx}

\newcommand{\Z}{{\mathbb{Z}}}

\newcommand{\pa}{\partial}

\newcommand{\vep}{\varepsilon}

\begin{document}

\title[Casimir pistons with mixed boundary conditions]{Finite temperature Casimir pistons for electromagnetic field with mixed boundary conditions and its classical limit}

\author{L.P. Teo}\email{lpteo@mmu.edu.my.}\affiliation{Faculty of Information
Technology, Multimedia University, Jalan Multimedia, Cyberjaya,
63100, Selangor Darul Ehsan, Malaysia.\\Phone: +6-03-83125344.}

\keywords{Casimir force, piston geometry, finite temperature, electromagnetic field,  mixed boundary conditions.
}
\pacs{11.10.Wx}

\begin{abstract}
In this paper, the finite temperature Casimir force acting on a two-dimensional Casimir piston due to electromagnetic field is computed. It was found that if mixed boundary conditions are assumed on  the piston and its opposite wall, then the Casimir force always tends to restore the piston towards the equilibrium position, regardless of the boundary conditions assumed on the walls transverse to the piston. In contrary, if pure boundary conditions are assumed on the piston and the opposite wall, then the Casimir force always tend to pull the piston towards the closer wall and away from the equilibrium position. The nature of the force is not affected by temperature. However, in the high temperature regime, the magnitude of the Casimir force grows linearly with respect to temperature. This shows that the Casimir effect has a classical limit as has been observed in other literatures.

\end{abstract}
\maketitle

\section{Introduction}
\begin{figure}\centering \epsfxsize=.5\linewidth
\epsffile{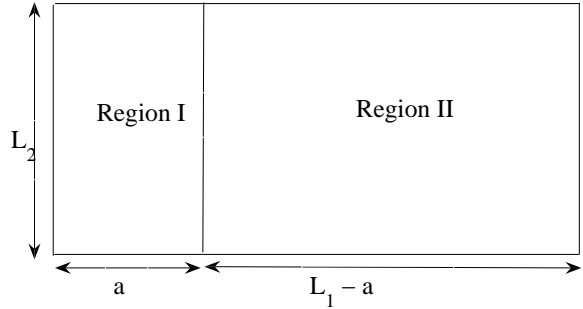} \caption{\label{Fig1} The two dimensional
rectangular piston}\end{figure}

Since the work of Cavalcanti \cite{1}, the Casimir effect of the piston geometry (see FiG. \ref{Fig1}) has attracted considerable interest for it was shown to be free of divergence problem. Some  studies have been  devoted to this subject \cite{13, 14, 15, 16, 17, 18, 19, 20, 21, 22, 23, 24, 25, 26, 27}. It was found that for massless scalar field with periodic boundary conditions (b.c.), Dirichlet b.c.~and Neumann b.c., and for electromagnetic field with perfect electric conductor (PEC) b.c.~and perfect magnetic conductor (PMC) b.c. in a $d$-dimensional space, the Casimir force acting on the piston always tends to pull the piston to the closest wall. This might create undesirable effect known as stiction in the functionality of nano devices. In \cite{16}, Barton showed that for a thin piston with weakly reflecting dielectrics, the Casimir force at small separations is attractive, but turn to repulsive  as the separation increases. Another scenario which brings to repulsive Casimir force     was considered in \cite{20, 27}, where a massless or massive scalar field  is assumed to satisfy Neumann b.c.~on the piston and   Dirichlet b.c. on all other walls. In this case, the zero temperature Casimir force was shown to be always repulsive. In \cite{21}, it was suggested that a perfectly conducting piston inside a rectangular cavity with infinitely permeable walls will lead to repulsive Casimir force.

In this paper, we   consider the thermal correction to the repulsive Casimir force due to electromagnetic  field with mixed boundary conditions (PEC b.c.~on one wall and   PMC b.c.~on the opposite wall) and determine whether temperature will change the nature of the force.
We only consider the case where the space dimension $d=2$. This will simplify the mathematical computation but it  gives enough indications for the general case of higher dimensions which will be considered in  future.  The two dimensional rectangular Casimir pistons for electromagnetic field with purely  PEC b.c.~and purely PMC b.c.~have been studied. The Casimir effect due to electromagnetic field with PMC b.c.~coincides with the Casimir effect due to a massless scalar field with Dirichlet b.c.~whose   zero temperature limit is studied in the pioneering work \cite{1}. The Casimir effect due to electromagnetic field with PEC b.c.~coincides with the Casimir effect due to a massless scalar field with Neumann b.c.~whose   zero temperature  limit is considered in \cite{19}. The finite temperature Casimir effect was recently considered in \cite{26}. It was found that for pure boundary conditions, the Casimir force is always attractive at any temperature. Therefore it will be interesting to see whether the thermal correction affects the repulsive nature of the Casimir force due to electromagnetic field with mixed b.c. This is the issue addressed in this paper. We consider more general case of mixed b.c.~where each pair of parallel plates can either assume pure boundary conditions (both PEC b.c.~or both PMC b.c.) or mixed boundary conditions.

In this paper, we work in the units where $\hbar$ (reduced Planck constant), $c$ (speed of light) and $k_B$ (Boltzmann constant) are equal to unity.
\section{Casimir energy for electromagnetic field with mixed boundary conditions inside a rectangular cavity}
Recall that    the finite temperature Casimir energy is defined as the sum of the zero temperature Casimir energy and the temperature correction, i.e.,
\begin{equation*}
E_{\text{Cas}}=E_{\text{Cas}}^0 + \Delta E_{\text{Cas}}= \frac{1}{2} \sum_{\omega_{\boldsymbol{k} }\neq 0}\omega_k + T\sum_{\omega_{\boldsymbol{k} }\neq 0}\log\left(1-e^{-\frac{\omega_{\boldsymbol{k} }}{T}}\right),
\end{equation*}where $\omega_{\boldsymbol{k}}$ runs through all zero point energies. The sum corresponding to the temperature correction $\Delta E_{\text{Cas}}$ is a convergent sum. However, the zero temperature contribution $E_{\text{Cas}}^0$ is divergent. There are different ways to regularize this sum. In the zeta regularization scheme \cite{5, 6, 7, 8}, we define the zeta function
\begin{equation*}
\zeta_0(s) =\sum_{\omega_{\boldsymbol{k} }\neq 0} \omega_{\boldsymbol{k} }^{-2s}
\end{equation*}and  analytically continue it to a neighborhood of $s=-1/2$. If $\zeta_0(s)$ is regular at $s=-1/2$, the zeta--regularized zero temperature Casimir energy is then defined as
\begin{equation*}
E_{\text{Cas}}^{0, \text{zeta reg}}=\frac{1}{2}\zeta_0\left(-\frac{1}{2}\right).
\end{equation*} Correspondingly, the finite temperature Casimir energy can be computed by using the zeta function
\begin{equation*}
\zeta(s) =\sum_{\omega_{\boldsymbol{k} }\neq 0} \sum_{l=-\infty}^{\infty} \left(\omega_{\boldsymbol{k} }^2+ (2\pi l T)^2\right)^{-s}.
\end{equation*}It can be shown that (see \cite{9, 10, 11, 12}) if $\zeta(s)$ has an analytic continuation to a neighborhood of $s=0$ with $\zeta(0)=0$, then
\begin{equation*}
\zeta'(0)= -\frac{1}{T}\zeta_0\left(-\frac{1}{2}\right)-2\sum_{\omega_k\neq 0}\log\left(1-e^{-\frac{\omega_{\boldsymbol{k} }}{T}}\right).
\end{equation*}Consequently, the zeta regularized finite temperature Casimir energy is equal to
\begin{equation*}
E_{\text{Cas}}^{ \text{  reg}}=-\frac{T}{2}\zeta'(0).
\end{equation*}A disadvantage of applying the zeta regularization scheme is that all the   divergence terms in the Casimir energy has been renormalized to zero. However, it can be shown as in \cite{26} that in the piston scenario, the divergence terms of the Casimir force acting on the piston due to Region I and Region II always cancel without renormalization due to the fact that the divergence terms of the Casimir energies are linear in $L_1$.

For an electromagnetic field inside a $d$-dimensional space $\Omega$, the field strength is represented by a totally anti-symmetric rank two tensor $\mathfrak{F}^{\mu\nu}$, $\mu,\nu=0,1,\ldots, d$, satisfying the equations
\begin{align}\label{eq8_14_1}
\pa_{\mu} \tilde{\mathfrak{F}}^{\mu \nu_{1}\ldots\nu_{d-2}}=0, \hspace{1cm}\pa_{\mu}\mathfrak{F}^{\mu\nu}=j^{\nu},
\end{align}where $\tilde{\mathfrak{F}}^{\mu_1\ldots\mu_{d-1}}=\vep^{\mu_1\ldots\mu_{d-1},\nu,\lambda}\mathfrak{F}_{\nu\lambda}$ is the dual tensor of $\mathfrak{F}^{\mu\nu}$ and $j^{\nu}$ is the current. In the vacuum state, $j^{\nu}=0$. There are two ideal boundary conditions that are of particular interest, i.e., the perfect electric conductor (PEC) boundary conditions (b.c.)~characterized by $n_{\mu}\left.\tilde{\mathfrak{F}}^{\mu\nu_1\ldots\nu_{d-2}} \right|_{\pa\Omega}=0$ and the perfect magnetic conductor (PMC) b.c.~characterized by $n_{\mu}\left. \mathfrak{F}^{\mu\nu} \right|_{\pa\Omega}=0$. Introducing the potentials $A^{\mu}$ so that
\begin{equation*}
F^{\mu\nu}=\pa^{\mu}A^{\nu}-\pa^{\nu}A^{\mu}, \hspace{0.5cm}\pa^0=\pa_0, \;\;\pa^i=-\pa_i, \;\;1\leq i\leq d;
\end{equation*}
and working in the radiation gauge $$A^0=0, \hspace{1cm} \pa_iA^i=0,$$ eq.~\eqref{eq8_14_1} is equivalent to
\begin{equation*}
\Delta A^i=0, \hspace{1cm}\Delta:=\pa_0^2-\sum_{j=1}^d \pa_j^2,
\end{equation*}when $j^{\mu}=0$. When the space $\Omega$ is a rectangular cavity $\Omega=[0, L_1]\times \ldots\times [0, L_d]$, the PEC b.c.~on a wall $x_i=0$ or $x_i=L_i$ is equivalent to $$\left.\pa_{\mu}A_{\nu} -\pa_{\nu}A_{\mu}\right|_{x_i=0\;\text{or}\;x_i=L_i}=0$$ for all $\mu \neq \nu \in\{0, 1, \ldots, d\}\setminus\{i\}$; whereas the PMC b.c.~is equivalent to $$\left.\pa_{i}A_{\mu}-\pa_{\mu}A_i\right|_{x_i=0\;\text{or}\;x_i=L_i}=0$$  for all $\mu  \in\{0, 1, \ldots, d\}\setminus\{i\}.$
Restricted to the case $d=2$, we consider the following different combinations of boundary conditions:

\vspace{0.2cm}
\noindent \textbf{Case I} Mixed boundary conditions (i.e., one wall PEC b.c.~and one wall PMC b.c.)~on both $x_1$ and $x_2$ directions.

\vspace{0.2cm}
\noindent \textbf{Case II} Mixed boundary conditions   on one direction, say $x_1$, and purely PEC b.c.~in the other direction.

\vspace{0.2cm}
\noindent \textbf{Case III} Mixed boundary conditions   on one direction, say $x_1$, and purely PMC b.c.~in the other direction.

\vspace{0.2cm} Now we derive the finite temperature Casimir energy of the electromagnetic field for each of the above boundary conditions:

\vspace{0.2cm}
\noindent \textbf{Case I} In this case, we are looking for solutions of $A_1(x_1, x_2, t)$ and $A_2(x_1, x_2, t)$ satisfying\begin{equation}\label{eq8_21_1}(\pa_t^2-\pa_{x_1}^2-\pa_{x_2}^2) A_i=0, \;\; i=1,2; \;\;\;\pa_{x_1}A_1+\pa_{x_2}A_2=0,\end{equation} and the boundary conditions
\begin{equation*} \begin{split}\left.\pa_t A_1\right|_{x_1=L_1, x_2=0}=0,\hspace{0.5cm}\left.\pa_t A_2\right|_{x_1=0, x_2=L_2}=0, \hspace{0.5cm}\left.\left(\pa_{x_1} A_2-\pa_{x_2}A_1\right)\right|_{x_1=L_1, x_2=L_2}=0.\end{split}\end{equation*} It is easy to verify that a basis of solutions are given by
\begin{equation*}
\begin{pmatrix} A_1(x_1, x_2, t)\\A_2(x_1, x_2, t)\end{pmatrix} = \begin{pmatrix}\alpha_1 \cos\frac{\pi\left(k_1+\frac{1}{2}\right)x_1}{L_1}\sin\frac{\pi\left(k_2+\frac{1}{2}\right)x_2}{L_2}\\ \alpha_2 \sin\frac{\pi\left(k_1+\frac{1}{2}\right)x_1}{L_1}\cos\frac{\pi\left(k_2+\frac{1}{2}\right)x_2}{L_2}\end{pmatrix}e^{-\omega_{\boldsymbol{k}}t},\;\;\;\;k_1, k_2\in \tilde{\mathbb{N}}=\mathbb{N}\cup\{0\},
\end{equation*} subjected to the condition
\begin{equation*}
\frac{\alpha_1\left(k_1+\frac{1}{2}\right)}{L_1}+\frac{\alpha_2\left(k_2+\frac{1}{2}\right)}{L_2}=0.
\end{equation*}Here   $$\omega_{\boldsymbol{k} } =\pi\sqrt{\left(\frac{k_1+\frac{1}{2}}{L_1}\right)^2+\left(\frac{k_2+\frac{1}{2}}{L_2}\right)^2} .$$
The corresponding zeta function is
\begin{equation*}\begin{split}
\zeta(s)= \frac{\pi^{-2s}}{4}\Biggl\{ Z_3\left(s; \frac{1}{2L_1},\frac{1}{2L_2}, 2T\right)-Z_3\left(s; \frac{1}{2L_1}, \frac{1}{L_2}, 2T\right)\\-Z_3\left(s; \frac{1}{L_1}, \frac{1}{2L_2}, 2T\right)+Z_3\left(s; \frac{1}{L_1},\frac{1}{L_2}, 2T\right)\Biggr\},
\end{split}\end{equation*}where $Z_n\left(s; c_1, \ldots, c_n\right)$ is the homogeneous Epstein zeta function defined by
\begin{equation}\label{eq8_26_1}
Z_n(s; c_1, \ldots, c_n) =\sum_{\boldsymbol{k}\in \widehat{\mathbb{Z}^n}} \left(\sum_{j=1}^n [c_j k_j]^2\right)^{-s},
\end{equation}and $\widehat{\mathbb{Z}^n}=\mathbb{Z}^n\setminus\{ \mathbf{0}\}$. Since $Z_n(0; c_1, \ldots, c_n)=-1$, we find that the regularized Casimir energy for electromagnetic field with mixed boundary conditions in both $x_1$ and $x_2$ directions of a rectangular cavity  is given by
\begin{equation}\label{eq8_27_4}\begin{split}
E_{\text{Cas}}^{\text{I, reg}}(L_1, L_2) =&- \frac{T}{8}\Biggl\{ Z_3'\left(0; \frac{1}{2L_1},\frac{1}{2L_2}, 2T\right)-Z_3'\left(0; \frac{1}{2L_1}, \frac{1}{L_2}, 2T\right)\\&-Z_3'\left(0; \frac{1}{L_1}, \frac{1}{2L_2}, 2T\right)+Z_3'\left(0; \frac{1}{L_1},\frac{1}{L_2}, 2T\right)\Biggr\}.\end{split}
\end{equation}Explicit formulas for $Z_n'(0; c_1, \ldots, c_n)$ are given in the Appendix \ref{appendix}.

\vspace{0.2cm}
\noindent \textbf{Case II} In this case, we are looking for solutions of $A_1(x_1, x_2, t)$ and $A_2(x_1, x_2, t)$ satisfying \eqref{eq8_21_1} and the boundary conditions
\begin{equation*} \begin{split}\left.\pa_t A_1\right|_{x_1=L_1, x_2=0, x_2=L_2}=0,\hspace{0.5cm}\left.\pa_t A_2\right|_{x_1=0 }=0, \hspace{0.5cm}\left.\left(\pa_{x_1} A_2-\pa_{x_2}A_1\right)\right|_{x_1=L_1 }=0.\end{split}\end{equation*}A basis of solutions are given by
\begin{equation*}
\begin{pmatrix} A_1(x_1, x_2, t)\\A_2(x_1, x_2, t)\end{pmatrix} = \begin{pmatrix}\alpha_1 \cos\frac{\pi\left(k_1+\frac{1}{2}\right)x_1}{L_1}\sin\frac{\pi k_2x_2}{L_2}\\ \alpha_2 \sin\frac{\pi\left(k_1+\frac{1}{2}\right)x_1}{L_1}\cos\frac{\pi k_2x_2}{L_2}\end{pmatrix}e^{-\omega_{\boldsymbol{k}} t}, \;\;\;\;  k_1, k_2\in \tilde{\mathbb{N}},\end{equation*}\begin{equation*}\omega_{\boldsymbol{k}}=\pi\sqrt{\left(\frac{k_1+\frac{1}{2}}{L_1}\right)^2 +\left(\frac{k_2}{L_2}\right)^2},
\end{equation*}subjected to the condition
\begin{equation*}
\frac{\alpha_1\left(k_1+\frac{1}{2}\right)}{L_1}+\frac{\alpha_2 k_2}{L_2}=0.
\end{equation*}The corresponding regularized Casimir energy is
\begin{equation}\label{eq8_27_5}\begin{split}
E_{\text{Cas}}^{\text{II, reg}}(L_1, L_2) =& -\frac{T}{8}\Biggl\{ Z_3'\left(0; \frac{1}{2L_1},\frac{1}{L_2}, 2T\right)-Z_3'\left(0; \frac{1}{L_1}, \frac{1}{L_2}, 2T\right)\\&+Z_2'\left(0; \frac{1}{2L_1},  2T\right)-Z_2'\left(0; \frac{1}{L_1}, 2T\right)\Biggr\}.\end{split}
\end{equation}

\vspace{0.2cm}
\noindent \textbf{Case III} In this case, we are looking for solutions of $A_1(x_1, x_2, t)$ and $A_2(x_1, x_2, t)$ satisfying \eqref{eq8_21_1} and the boundary conditions
\begin{equation*}\begin{split} \left.\pa_t A_1\right|_{x_1=L_1 }=0,\hspace{0.5cm}\left.\pa_t A_2\right|_{x_1=0, x_2=0, x_2=L_2 }=0, \hspace{0.5cm}\left.\left(\pa_{x_1} A_2-\pa_{x_2}A_1\right)\right|_{x_1=L_1, x_2=0, x_2=L_2 }=0.\end{split}\end{equation*} A basis of solutions are given by
\begin{equation*}
\begin{pmatrix} A_1(x_1, x_2, t)\\A_2(x_1, x_2, t)\end{pmatrix} = \begin{pmatrix}\alpha_1 \cos\frac{\pi \left(k_1+\frac{1}{2}\right)x_1}{L_1}\cos\frac{\pi k_2x_2}{L_2}\\ \alpha_2 \sin\frac{\pi \left(k_1+\frac{1}{2}\right)x_1}{L_1}\sin\frac{\pi k_2x_2}{L_2}\end{pmatrix}e^{-\omega_{\boldsymbol{k}} t},\;\;\;\;k_1\in \tilde{\mathbb{N}}, k_2\in\mathbb{N},
\end{equation*}\begin{equation*}\omega_{\boldsymbol{k}}=\pi\sqrt{\left(\frac{k_1+\frac{1}{2}}{L_1}\right)^2 +\left(\frac{k_2}{L_2}\right)^2},
\end{equation*}subjected to the condition
\begin{equation*}
-\frac{\alpha_1\left(k_1+\frac{1}{2}\right)}{L_1}+\frac{\alpha_2 k_2}{L_2}=0.
\end{equation*}
The corresponding regularized Casimir energy is
\begin{equation}\label{eq8_27_6}\begin{split}
E_{\text{Cas}}^{\text{III, reg}}(L_1, L_2) =& -\frac{T}{8}\Biggl\{ Z_3'\left(0; \frac{1}{2L_1},\frac{1}{L_2}, 2T\right)-Z_3'\left(0; \frac{1}{L_1}, \frac{1}{L_2}, 2T\right)\\&-Z_2'\left(0; \frac{1}{2L_1},  2T\right)+Z_2'\left(0; \frac{1}{L_1}, 2T\right)\Biggr\}.\end{split}
\end{equation}

Notice that there is a slight difference between the set of eigenmodes  in Case II and Case III. In Case II, we allow $k_2=0$ which corresponds to solutions \begin{equation*}
\begin{pmatrix} A_1(x_1, x_2, t)\\A_2(x_1, x_2, t)\end{pmatrix} = \begin{pmatrix}0\\ \alpha_2 \sin\frac{\pi\left(k_1+\frac{1}{2}\right)x_1}{L_1} \end{pmatrix}e^{-\omega_k t},\hspace{1cm} k_1\in\tilde{\mathbb{N}}, \;\;\omega_k=\frac{\pi\left(k_1+\frac{1}{2}\right)}{L_1}.\end{equation*}
However, in Case III, $k_2=0$ implies that $\alpha_1=0$ and $A_1=A_2\equiv 0$. Therefore there is no eigenmode with $k_2=0$.

\section{Casimir force acting on the piston for electromagnetic field with mixed boundary conditions}

In this section, we consider the Casimir force acting on a  two-dimensional rectangular piston due to electromagnetic field with mixed boundary conditions. The boundary conditions on the walls of Region I are the Cases I, II, III as considered in the previous section. In Region II, we assume that the boundary condition on the wall $x_1=L_1$ is the same as on the wall $x_1=0$. We have the following cases.

\subsection{Case MBC-A} We assume mixed boundary conditions on both directions. In this case, we find that the total regularized Casimir energy of the piston system is
\begin{equation*}
E_{\text{Cas}}^{A, \text{reg}}(a; L_1, L_2)=E_{\text{Cas}}^{\text{I, reg}}(a, L_2)+E_{\text{Cas}}^{\text{I, reg}}(L_1-a, L_2).
\end{equation*}Applying Chowla--Selberg formula \eqref{eq8_27_3} to \eqref{eq8_27_4}, we find that
\begin{equation*}\begin{split}
E_{\text{Cas}}^{\text{I, reg}}(L_1, L_2)=&-\frac{T}{8} \Biggl\{ \frac{L_1L_2}{2\pi T}Z_2\left(\frac{3}{2}; 2L_2,\frac{1}{2T}\right)-\frac{L_1L_2}{4\pi T} Z_2\left(\frac{3}{2}; L_2, \frac{1}{2T}\right)
\\&+4\sum_{k_1=1}^{\infty}\sum_{k_2=0}^{\infty} \sum_{l=-\infty}^{\infty} \frac{(-1)^{k_1}}{k_1}  \exp\left(-2\pi k_1L_1\sqrt{\left(\frac{k_2+\frac{1}{2}}{L_2}\right)^2+(2lT)^2}\right)\Biggr\}.
\end{split}
\end{equation*}Therefore, in the limit $L_1\rightarrow \infty$, the Casimir force acting on the piston is given by
\begin{equation}\label{eq8_27_7}\begin{split}
F_{\text{Cas}}^{\text{A}, L_1=\infty}(a;   L_2)=&\lim_{L_1\rightarrow \infty}F_{\text{Cas}}^{\text{A}}(a; L_1, L_2)\\=&-\lim_{L_1\rightarrow \infty}\frac{\pa}{\pa a}E_{\text{Cas}}^{A, \text{reg}}(a; L_1, L_2)\\
=&\pi T \sum_{k_2=0}^{\infty} \sum_{l=-\infty}^{\infty}  \frac{\sqrt{\left(\frac{k_2+\frac{1}{2}}{L_2}\right)^2+(2lT)^2}}{\exp\left(2\pi a\sqrt{\left(\frac{k_2+\frac{1}{2}}{L_2}\right)^2+(2lT)^2}\right)+1}.
\end{split}\end{equation}Notice that this is a positive decreasing function in $a$. Consequently, when $L_1$ is finite, the Casimir force acting on the piston
\begin{equation*}
F_{\text{Cas}}^{\text{A}}(a; L_1, L_2)=F_{\text{Cas}}^{\text{A}, L_1=\infty}(a;   L_2)-F_{\text{Cas}}^{\text{A}, L_1=\infty}(L_1-a;   L_2)
\end{equation*}is positive if $a<L_1-a$, and is negative if $a>L_1-a$. In other words, at any temperature, the Casimir force always tends to restore the piston to the equilibrium position $x_1=L_1/2$, which is the middle of the cavity.

The infinite summation in the expression \eqref{eq8_27_7} for the Casimir force   converges very fast if $a\gg L_2$. It shows that in the limit $L_1\rightarrow \infty$,   the magnitude of the Casimir force decays exponentially when the plate separation $a$ is large. In most practical situation, we are interested in the opposite case where $a\ll L_2$. In this latter case, the Chowla--Selberg formula \eqref{eq8_27_3} gives
\begin{equation}\label{eq8_28_2}
\begin{split}
&F_{\text{Cas}}^{\text{A}, L_1=\infty}(a;   L_2)=\frac{3\zeta_R(3)}{32\pi }\frac{L_2}{a^3} -\frac{L_2}{32\pi}\sum_{(k_3,\ell)\in \widehat{\Z^2}}
\frac{(-1)^{k_2}}{\left([k_2L_2]^2+\left[\frac{l}{2T}\right]^2\right)^{\frac{3}{2}}}
\\&+\frac{\pi L_2}{2a^3} \sum_{k_1=0}^{\infty}\sum_{(k_2,\ell)\in\widehat{\Z^2}}  (-1)^{k_2} \left(k_1+\frac{1}{2}\right)^2  K_0\left(\frac{2\pi\left(k_1+\frac{1}{2}\right)}{a}
\sqrt{[k_2L_2]^2+\left[\frac{l}{2T}\right]^2}\right).
\end{split}
\end{equation}This shows that at any temperature, when the plate separation $a$ is small, the leading behavior of the Casimir force is given by
\begin{equation*}
F_{\text{Cas}}^{\text{A}, L_1=\infty}(a;   L_2)\sim \frac{3\zeta_R(3)}{32\pi }\frac{L_2}{a^3} +O(a^0).
\end{equation*}It implies that when $a\rightarrow 0^+$, the magnitude of the Casimir force approaches $\infty$ and behaves as $1/a^3$. From this we can conclude that at any temperature,   the Casimir force acting on the piston, considered  as a function of $a\in (0, L_1)$, decreases from $\infty$ to $0$ when $a\in (0, L_1/2)$ and increases from $0$ to $\infty$ when $a\in (L_1/2, L_1)$.

The formula \eqref{eq8_27_7} can also be used to study the high temperature behavior of the Casimir force. It shows that in the high temperature regime, the leading behavior of the Casimir force is
\begin{equation}\label{eq8_28_1}
\begin{split}
F_{\text{Cas}}^{\text{A}}(a; L_1, L_2)\sim &\frac{\pi T}{L_2} \sum_{k_2=0}^{\infty}    \frac{ k_2+\frac{1}{2} }{\exp\left(\frac{2\pi a}{L_2}\left(k_2+\frac{1}{2} \right)\right)+1} -\left(a\longleftrightarrow L_1-a\right),
\end{split}\end{equation}which is linear in $T$. The remaining terms decays exponentially as $T\rightarrow \infty$. If we restore the units $\hbar$, $c$ and $k_B$ into the expression for Casimir force, we find that a term with $T^j$ will be accompanied by $\hbar^{j-1}$. Therefore \eqref{eq8_28_1} shows that the Casimir force acting on the piston has a classical ($\hbar\rightarrow 0$) limit, as has also been observed in other works in Casimir effect (see e.g. \cite{1_15_1, 1_15_2, 1_15_3, 1_15_4}). The right hand side of \eqref{eq8_28_1} is called the classical term of the Casimir force.

  In the low temperature ($T\ll 1$) regime, the Casimir force is dominated by the zero temperature Casimir force, with correction term being the temperature correction:
\begin{equation*}
F_{\text{Cas}}^{\text{A}}(a; L_1, L_2)=F_{\text{Cas}}^{\text{A}, T=0}(a; L_1, L_2)+\Delta_T F_{\text{Cas}}^{\text{A}}(a; L_1, L_2).
\end{equation*} Applying the Chowla--Selberg formula \eqref{eq8_27_1}, we have
\begin{equation*}\begin{split}
-\frac{L_2}{32\pi}\sum_{(k_3,\ell)\in \widehat{\Z^2}}
\frac{(-1)^{k_2}}{\left([k_2L_2]^2+\left[\frac{l}{2T}\right]^2\right)^{\frac{3}{2}}}
=&-\frac{L_2}{32\pi}\left( 2Z_2\left(\frac{3}{2}; 2L_2,\frac{1}{2T}\right)-Z_2\left(\frac{3}{2}; L_2,\frac{1}{2T}\right)\right)\\
=&\frac{3\zeta_R(3)}{64\pi L_2^2}- \frac{T}{ L_2}\sum_{k_2=0}^{\infty}\sum_{l=1}^{\infty} \frac{k_2+\frac{1}{2}}{l}K_1\left(\frac{\pi \left(k_2+\frac{1}{2}\right)l}{L_2T}\right).
\end{split}
\end{equation*}With this, we can read from the formula \eqref{eq8_28_2} that the zero temperature Casimir force is given by
\begin{equation*}\begin{split}
F_{\text{Cas}}^{\text{A}, T=0}(a; L_1, L_2)=&\frac{3\zeta_R(3)}{32\pi }\frac{L_2}{a^3} +\frac{3\zeta_R(3)}{64\pi L_2^2}
+\frac{\pi L_2}{a^3} \sum_{k_1=0}^{\infty}\sum_{k_2=1}^{\infty} (-1)^{k_2} \left(k_1+\frac{1}{2}\right)^2  K_0\left(\frac{2\pi k_2\left(k_1+\frac{1}{2}\right)L_2}{a}\right)\\
&-\left(a\longleftrightarrow L_1-a\right);\end{split}
\end{equation*}and the thermal correction is
\begin{equation*}\begin{split}
&\Delta_T F_{\text{Cas}}^{\text{A}}(a; L_1, L_2)=- \frac{T}{ L_2}\sum_{k_2=0}^{\infty}\sum_{l=1}^{\infty} \frac{k_2+\frac{1}{2}}{l}K_1\left(\frac{\pi \left(k_2+\frac{1}{2}\right)l}{L_2T}\right)+\frac{\pi L_2}{a^3} \sum_{k_1=0}^{\infty}\sum_{k_2=-\infty}^{\infty}\sum_{l=1}^{\infty}\\&\times (-1)^{k_2} \left(k_1+\frac{1}{2}\right)^2    K_0\left(\frac{2\pi\left(k_1+\frac{1}{2}\right)}{a}
\sqrt{[k_2L_2]^2+\left[\frac{l}{2T}\right]^2}\right) -\left(a\leftrightarrow L_1-a\right).
\end{split}
\end{equation*}Notice that if $L_1\rightarrow \infty$, the thermal correction to the Casimir force decays to zero exponentially fast when $T\rightarrow 0^+$.

In the limit $L_1, L_2\rightarrow \infty$, the geometric configuration becomes that of a pair of infinite parallel plates separated by a distance $a$. In this case, since
\begin{equation*}\begin{split}
-\frac{L_2}{32\pi}\sum_{(k_3,\ell)\in \widehat{\Z^2}}
\frac{(-1)^{k_2}}{\left([k_2L_2]^2+\left[\frac{l}{2T}\right]^2\right)^{\frac{3}{2}}}
=-\frac{L_2T^3}{2\pi}\zeta_R(3) +\frac{\pi}{48}\frac{T}{L_2} -2T^2\sum_{k_2=1}^{\infty}\sum_{l=1}^{\infty}\frac{(-1)^{k_2}}{k_2} l K_1(4\pi k_2 l L_2T),
\end{split}
\end{equation*}eq.~\eqref{eq8_28_2} then implies that in the infinite parallel plates limit, the Casimir force acting on a wall is given by
\begin{equation}\label{eq9_2_1}\begin{split}
F_{\text{Cas}}^{\text{A}, ||} (a)=&L_2\Biggl\{ \frac{3\zeta_R(3)}{32\pi a^3 }-\frac{T^3}{2\pi}\zeta_R(3)
+\frac{\pi}{a^3}\sum_{k_1=0}^{\infty}\sum_{l=1}^{\infty} \left(k_1+\frac{1}{2}\right)^2K_0\left(\frac{\pi l\left(k_1+\frac{1}{2}\right)}{a T}\right)\Biggr\}.
\end{split}\end{equation}This shows that for infinite parallel plates, the zero temperature Casimir force is
\begin{equation*}
F_{\text{Cas}}^{\text{A}, ||, T=0}(a) = \frac{3\zeta_R(3)}{32\pi a^3 }L_2.
\end{equation*}The temperature correction is of order $T^3$ as $T\rightarrow 0^+$. The remaining terms decays to zero exponentially fast when $T\rightarrow 0^+$. In the high temperature regime,
\begin{equation*}\begin{split}
F_{\text{Cas}}^{\text{A}, || }(a) =& L_2\Biggl\{ \frac{\pi }{48 a^2}T -\frac{2T^2}{a}\sum_{k_1=1}^{\infty}\sum_{l=1}^{\infty} (-1)^{k_1} \frac{l}{k_1} K_1(4\pi l k_1 Ta) -8\pi T^3\sum_{k_1=1}^{\infty}\sum_{l=1}^{\infty} (-1)^{k_1} l^2 K_0(4\pi l k_1 Ta)\Biggr\}.\end{split}
\end{equation*}This shows that the classical limit of the Casimir force acting on a pair of infinite parallel plates with mixed boundary conditions is $$F_{\text{Cas}}^{\text{A}, ||, \text{classical} }(a) =  \frac{\pi L_2 }{48 a^2}T.$$

\subsection{Case MBC-B} We assume mixed boundary conditions in the $x_1$ direction and purely PEC b.c. in $x_2$ direction. Using the same method as the previous section, we find that the Casimir force acting on the piston is given by
\begin{equation*}
F_{\text{Cas}}^{\text{B}}(a; L_1, L_2)=F_{\text{Cas}}^{\text{B}, L_1=\infty}(a;   L_2)-F_{\text{Cas}}^{\text{B}, L_1=\infty}(L_1-a;   L_2),
\end{equation*}where
\begin{equation}\label{eq8_28_4}\begin{split}&F_{\text{Cas}}^{\text{B}, L_1=\infty}(a;   L_2) =\pi T \sum_{(k_2,  l)\in \tilde{\mathbb{N}}\times \Z \setminus\{\mathbf{0}\}}  \frac{\sqrt{\left(\frac{k_2 }{L_2}\right)^2+(2lT)^2}}{\exp\left(2\pi a\sqrt{\left(\frac{k_2 }{L_2}\right)^2+(2lT)^2}\right)+1}.
\end{split}
\end{equation}As in the previous case, this shows that at any temperature, the Casimir force tends to pull the piston to the equilibrium position $x_1=L_1/2$.  Moreover, it shows that in the high temperature limit, the leading term of the Casimir force is given by the classical term
\begin{equation}\label{eq8_28_6}\begin{split}F_{\text{Cas}}^{\text{B}}(a; L_1, L_2)\sim &\frac{\pi T}{L_2} \sum_{k_2=1}^{\infty}  \frac{ k_2 }{\exp\left(\frac{2\pi k_2 a}{L_2}\right)+1}-\left(a\leftrightarrow L_1-a\right).
\end{split}\end{equation}

An alternative expression for $F_{\text{Cas}}^{\text{B}, L_1=\infty}(a;   L_2)$ that can be used to study the small $a$ and low $T$ behavior of the Casimir force is
\begin{equation}\label{eq8_28_3}\begin{split}F_{\text{Cas}}^{\text{B}, L_1=\infty}(a;   L_2)=&\frac{3\zeta_R(3)}{32\pi }\frac{L_2}{a^3}+\frac{\pi}{96 a^2} -\frac{\zeta_R(3)}{16 \pi L_2^2}-\frac{\pi T^2}{6} -\frac{T}{L_2}\sum_{k_2=1}^{\infty}\sum_{l=1}^{\infty}\frac{k_2}{l}K_1\left(\frac{\pi k_2 l}{L_2 T}\right)
\\&+\frac{\pi L_2}{2a^3} \sum_{k_1=0}^{\infty}\sum_{(k_2,\ell)\in\widehat{\Z^2}}    \left(k_1+\frac{1}{2}\right)^2  K_0\left(\frac{2\pi\left(k_1+\frac{1}{2}\right)}{a}
\sqrt{[k_2L_2]^2+\left[\frac{l}{2T}\right]^2}\right)\\& +\frac{\pi}{2a^2}\sum_{k_1=0}^{\infty} \frac{\left(k_1+\frac{1}{2}\right)}{ \exp\left(\frac{\pi   \left(k_1+\frac{1}{2}\right)}{T a}\right)-1}.
\end{split}
\end{equation}It shows that when the plate separation $a$ is small, the leading terms of the Casimir force is given by
\begin{equation*}
F_{\text{Cas}}^{\text{B}}(a; L_1, L_2) \sim \frac{3\zeta_R(3)}{32\pi }\frac{L_2}{a^3}+\frac{\pi}{96 a^2} +O(a^0).
\end{equation*}Notice that the first term behaves as $1/a^3$ when $a\rightarrow 0^+$. On the other hand, \eqref{eq8_28_3} gives the zero temperature Casimir force as
\begin{equation} \begin{split}F_{\text{Cas}}^{\text{B}, T=0}(a; L_1,   L_2)=&\frac{3\zeta_R(3)}{32\pi }\frac{L_2}{a^3}+\frac{\pi}{96 a^2} -\frac{\zeta_R(3)}{16\pi  L_2^2} +\frac{\pi L_2}{a^3} \sum_{k_1=0}^{\infty}\sum_{k_2=1}^{\infty}   \left(k_1+\frac{1}{2}\right)^2  K_0\left(\frac{2\pi\left(k_1+\frac{1}{2}\right)k_2L_2}{a}
 \right) \\&-\left( a\longleftrightarrow L_1-a\right)
\end{split}
\end{equation}The thermal correction goes to zero exponentially fast when $T\rightarrow 0^+$.

In the parallel plate limit, it can be checked that one would obtain the same result as \eqref{eq9_2_1}. This should be expected since in the limit $L_2\rightarrow \infty$, the boundary conditions assumed on the $x_2$ direction become immaterial.
\subsection{Case MBC-C} We assume mixed boundary conditions in the $x_1$ direction and purely PMC b.c. in $x_2$ direction. This case is very similar to the MBC-B case. We find that the Casimir force acting on the piston is given by
\begin{equation*}
F_{\text{Cas}}^{\text{C}}(a; L_1, L_2)=F_{\text{Cas}}^{\text{C}, L_1=\infty}(a;   L_2)-F_{\text{Cas}}^{\text{C}, L_1=\infty}(L_1-a;   L_2),
\end{equation*}where
\begin{equation}\label{eq8_28_5}\begin{split}&F_{\text{Cas}}^{\text{C}, L_1=\infty}(a;   L_2)=\pi T\sum_{k_2=1}^{\infty}\sum_{l=-\infty}^{\infty} \frac{\sqrt{\left(\frac{k_2 }{L_2}\right)^2+(2lT)^2}}{\exp\left(2\pi a\sqrt{\left(\frac{k_2 }{L_2}\right)^2+(2lT)^2}\right)+1}
\end{split}
\end{equation}The difference between this term and the corresponding term in the case of MBC-B  lies in the summation over $k_2$, where now $k_2$ starts from 1 instead of $0$. As in the previous case, \eqref{eq8_28_5} shows that at any temperature, the Casimir force tends to pull the piston to the equilibrium position $x_1=L_1/2$.  Moreover, it shows that in the high temperature limit, the leading term of the Casimir force is given by the classical term
\begin{equation*}\begin{split}F_{\text{Cas}}^{\text{C}}(a; L_1, L_2)\sim &\frac{\pi T}{L_2} \sum_{k_2=1}^{\infty}  \frac{ k_2 }{\exp\left(\frac{2\pi k_2 a}{L_2}\right)+1}-\left(a\leftrightarrow L_1-a\right).
\end{split}\end{equation*}One notice that this classical term is the same as in the case of MBC-B given by \eqref{eq8_28_6}. In other words, the difference between the Casimir forces for case MBC-B and case MBC-C is insignificant at high temperature.

An alternative expression for $F_{\text{Cas}}^{\text{C}, L_1=\infty}(a;   L_2)$ that can be used to study the small $a$ and low $T$ behavior of the Casimir force is
\begin{equation}\label{eq8_28_7}\begin{split}F_{\text{Cas}}^{\text{C}, L_1=\infty}(a;   L_2)=&\frac{3\zeta_R(3)}{32\pi }\frac{L_2}{a^3}-\frac{\pi}{96 a^2} -\frac{\zeta_R(3)}{16 \pi L_2^2}  -\frac{T}{L_2}\sum_{k_2=1}^{\infty}\sum_{l=1}^{\infty}\frac{k_2}{l}K_1\left(\frac{\pi k_2 l}{L_2 T}\right)
\\&+\frac{\pi L_2}{2a^3} \sum_{k_1=0}^{\infty}\sum_{(k_2,\ell)\in\widehat{\Z^2}}    \left(k_1+\frac{1}{2}\right)^2  K_0\left(\frac{2\pi\left(k_1+\frac{1}{2}\right)}{a}
\sqrt{[k_2L_2]^2+\left[\frac{l}{2T}\right]^2}\right)\\& -\frac{\pi}{2a^2}\sum_{k_1=0}^{\infty} \frac{\left(k_1+\frac{1}{2}\right)}{ \exp\left(\frac{\pi   \left(k_1+\frac{1}{2}\right)}{T a}\right)-1}.
\end{split}
\end{equation}When the plate separation $a$ is small, the leading terms of the Casimir force is given by
\begin{equation*}
F_{\text{Cas}}^{\text{C}}(a; L_1, L_2) \sim \frac{3\zeta_R(3)}{32\pi }\frac{L_2}{a^3}-\frac{\pi}{96 a^2} +O(a^0),
\end{equation*}with leading order $1/a^3$ when $a\rightarrow 0^+$. On the other hand,  the zero temperature Casimir force is
\begin{equation} \label{eq9_3_1}\begin{split}F_{\text{Cas}}^{\text{C}, T=0}(a; L_1,   L_2)=&\frac{3\zeta_R(3)}{32\pi }\frac{L_2}{a^3}-\frac{\pi}{96 a^2} -\frac{\zeta_R(3)}{16 \pi L_2^2} +\frac{\pi L_2}{a^3} \sum_{k_1=0}^{\infty}\sum_{k_2=1}^{\infty}   \left(k_1+\frac{1}{2}\right)^2  K_0\left(\frac{2\pi\left(k_1+\frac{1}{2}\right)k_2L_2}{a}
 \right)\\&-\left( a\longleftrightarrow L_1-a\right),
\end{split}
\end{equation}which only differs with the MBC-B case by the sign of the term $\pi/(96 a^2)$. The thermal correction also goes to zero exponentially fast when $T\rightarrow 0^+$.

We would   like to remark that the regularized Casimir energy and Casimir force acting on the piston in this case is the same as the corresponding quantities for massless scalar field which assume Neumann boundary condition on the piston and Dirichlet boundary conditions on the other walls. In fact, the zero temperature Casimir force \eqref{eq9_3_1} agrees with the corresponding result in \cite{20}.

\subsection{Case MBC-D} We assume PEC b.c. in the $x_1$ direction and mixed boundary conditions in $x_2$ direction. In this case,
\begin{equation*}
E_{\text{Cas}}^{D, \text{reg}}(a; L_1, L_2)=E_{\text{Cas}}^{\text{II, reg}}(L_2, a)+E_{\text{Cas}}^{\text{II, reg}}(L_2, L_1-a).
\end{equation*}
Similar computations give
\begin{equation*}
F_{\text{Cas}}^{\text{D}}(a; L_1, L_2)=F_{\text{Cas}}^{\text{D}, L_1=\infty}(a;   L_2)-F_{\text{Cas}}^{\text{D}, L_1=\infty}(L_1-a;   L_2),
\end{equation*}where
\begin{equation}\label{eq8_28_9}\begin{split}F_{\text{Cas}}^{\text{D}, L_1=\infty}(a;   L_2) =&- \pi T  \sum_{k_2=0}^{\infty}\sum_{l=-\infty}^{\infty}  \frac{\sqrt{\left(\frac{k_2 +\frac{1}{2}}{L_2}\right)^2+(2lT)^2}}{\exp\left(2\pi a\sqrt{\left(\frac{k_2+\frac{1}{2} }{L_2}\right)^2+(2lT)^2}\right)-1}.
\end{split}
\end{equation}Contrary to the previous cases, now we find that the Casimir force acting on the piston always tends to pull the piston towards the closer wall, and away from the equilibrium position. \eqref{eq8_28_9} also shows that in the high temperature regime, the Casimir force is dominated by the classical term, i.e.
\begin{equation*}\begin{split}
F_{\text{Cas}}^{\text{D}}(a; L_1, L_2) \sim &-  \frac{\pi T}{L_2} \sum_{k_2=0}^{\infty}  \frac{ k_2 +\frac{1}{2} }{\exp\left(\frac{2\pi a \left(k_2+\frac{1}{2} \right)}{L_2} \right)-1} - \left(a \leftrightarrow L_1-a\right)\end{split}
\end{equation*}as $T\rightarrow \infty$. The remaining terms decay exponentially.

An alternative expression for the Casimir force is given by
\begin{equation}\label{eq8_28_10}\begin{split}&F_{\text{Cas}}^{\text{D}}(a; L_1, L_2)=-\frac{L_2}{8\pi a^3}\zeta_R(3)+\frac{3\zeta_R(3)}{64\pi L_2^2}  - \frac{T}{ L_2}\sum_{k_2=0}^{\infty}\sum_{l=1}^{\infty}\frac{k_2+\frac{1}{2}}{l}K_1\left(\frac{\pi \left(k_2+\frac{1}{2}\right)l}{L_2T}\right)\\&+\frac{\pi L_2}{2a^3} \times\sum_{k_1=1}^{\infty}\sum_{(k_2,l)\in \widehat{Z^2}}(-1)^{k_2} k_1^2K_0\left( \frac{2\pi k_1}{a}\sqrt{(k_2L_2)^2+\left(\frac{l}{2T}\right)^2}\right) -\left(a\leftrightarrow L_1-a\right).
\end{split}
\end{equation}This shows that when the plate separation is small, the leading term of the Casimir force is
\begin{equation*}
F_{\text{Cas}}^{\text{D}}(a; L_1, L_2)\sim -\frac{L_2}{8\pi a^3}\zeta_R(3)+O(a^0),
\end{equation*}which is of order $1/a^3$. \eqref{eq8_28_10} also shows that in the low temperature limit, the Casimir force is dominated by the zero temperature Casimir force given by
\begin{equation*}\begin{split}
F_{\text{Cas}}^{\text{D}, T=0}(a; L_1, L_2)=&-\frac{L_2}{8\pi a^3}\zeta_R(3)+\frac{3\zeta_R(3)}{64\pi L_2^2}  +\frac{\pi L_2}{a^3} \sum_{k_1=1}^{\infty}\sum_{k_2=1}^{\infty}(-1)^{k_2} k_1^2K_0\left( \frac{2\pi k_1 k_2 L_2}{a} \right)\\& -\left(a\leftrightarrow L_1-a\right).\end{split}
\end{equation*} The thermal correction terms tends to zero exponentially fast when $T\rightarrow 0^+$.

In the infinite parallel plate limit, we find that
\begin{equation}\label{eq9_3_2}\begin{split}
F_{\text{Cas}}^{\text{D}, ||}(a)=&L_2\Biggl\{-\frac{\zeta_R(3)}{8\pi a^3}-\frac{ T^3}{2\pi} \zeta_R(3)+\frac{\pi }{a^3}\sum_{k_1=1}^{\infty}\sum_{l=1}^{\infty} k_1^2K_0\left(\frac{\pi l k_1}{aT}\right)\Biggr\},\end{split}
\end{equation}which gives the zero temperature Casimir force as
\begin{equation*}
F_{\text{Cas}}^{\text{D}, ||}(a)=-\frac{\zeta_R(3)L_2}{8\pi a^3},
\end{equation*}agreeing with well-known results (see e.g.~\cite{30}). An alternative expression for \eqref{eq9_3_2} is given by
\begin{equation*}\begin{split}
F_{\text{Cas}}^{\text{D}, ||}(a)=&L_2\Biggl\{-\frac{\pi T}{24 a^2}-\frac{2T^2}{a}\sum_{k_1=1}^{\infty}\sum_{l=1}^{\infty}\frac{l}{k_1}K_1(4\pi l k_1 Ta) -8\pi T^3\sum_{k_1=1}^{\infty}\sum_{l=1}^{\infty}l^2K_0(4\pi l k_1 Ta)\Biggr\},\end{split}
\end{equation*}which shows that the classical limit of the Casimir force is given by
\begin{equation*}
F_{\text{Cas}}^{\text{D}, ||, \text{classical}}(a)=-\frac{\pi  L_2}{24 a^2} T.
\end{equation*}

\subsection{Case MBC-E} We assume PMC b.c.~on $x_1$ direction and mixed boundary conditions on $x_2$ direction. In this case, although the regularized Casimir energy is different with the regularized Casimir energy for Case MBC-D, one can verify that their difference is a term independent of $L_1$. Consequently, the Casimir force acting on the piston for case MBC-E is identical to that for case MBC-D.

We do not discuss the cases where the electromagnetic field assumes purely PEC b.c.~on both directions or assumes purely PMC b.c.~on both directions. This has been considered in \cite{26}.
Another case we do not consider here is the case where the field assumes purely PEC b.c.~on one direction and purely PMC b.c.~on the other direction. The result is not much different from the cases of purely PEC b.c.~or purely PMC b.c.~on all directions.

\section{Discussion and conclusion} We have computed the exact formulas for the finite temperature Casimir force  acting on a two dimensional rectangular piston due to electromagnetic field with different combinations of boundary conditions. From the results, we can conclude that if mixed boundary conditions is assumed on the piston and its opposite wall, then the Casimir force always tend to move the piston to the equilibrium position, regardless of the boundary conditions assumed on the perpendicular walls. In contrary, if purely PMC b.c.~or purely PEC b.c.~is assumed on the piston and its opposite wall, then the Casimir force always tends to move the piston towards the closer wall, again regardless of the boundary conditions assumed on the perpendicular walls. This nature of the force is not affected by the change of temperature.
However, as in the case of pure boundary conditions   discussed in \cite{26}, the magnitude of the Casimir force grows linearly with temperature when the temperature is high enough. The implication of this is that although Casimir force is a quantum effect, it has a classical limit, as has been observed in \cite{1_15_1, 1_15_2, 1_15_3, 1_15_4}.

\begin{figure} \epsfxsize=0.33\linewidth
\epsffile{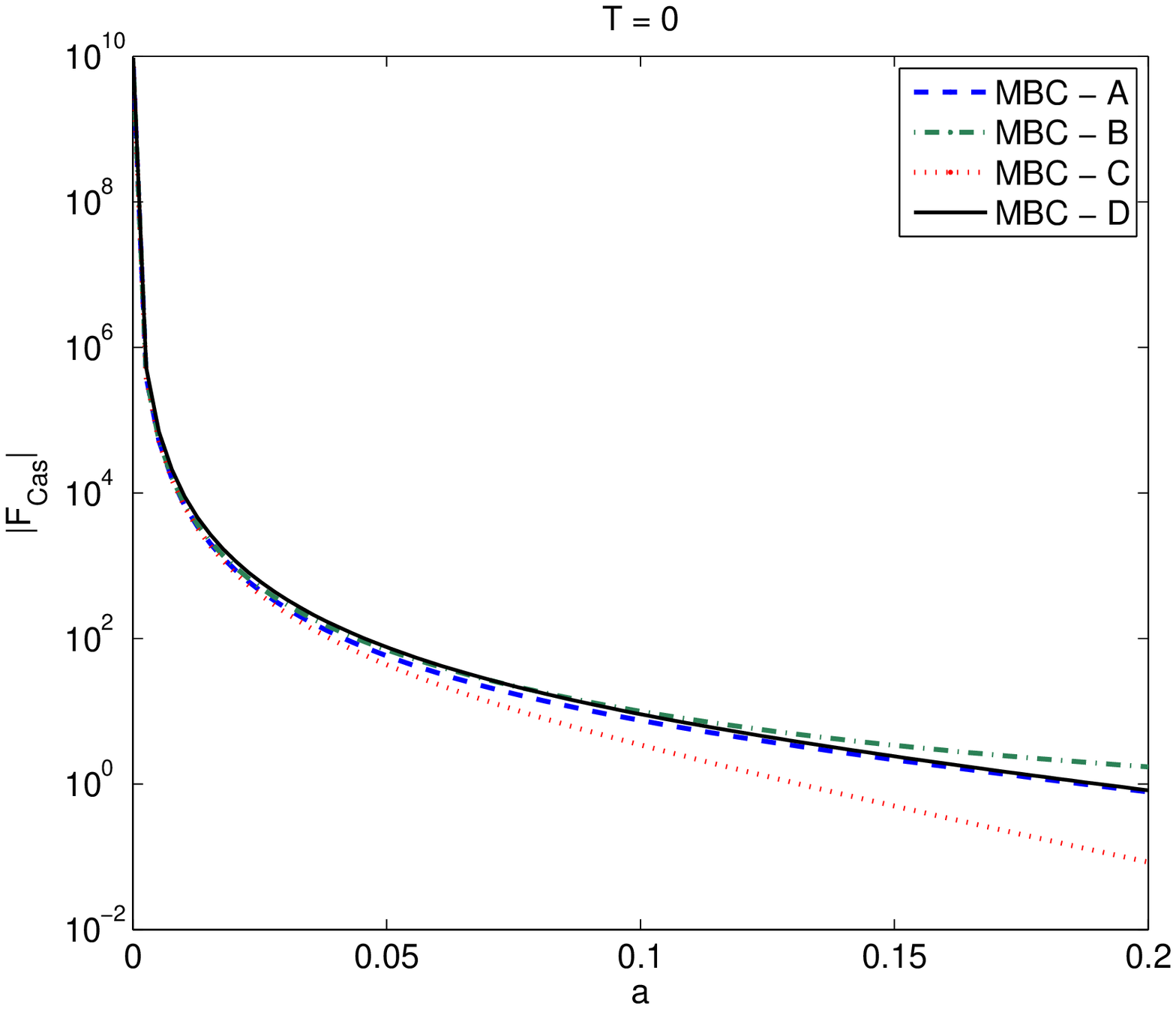}\epsfxsize=0.33\linewidth
\epsffile{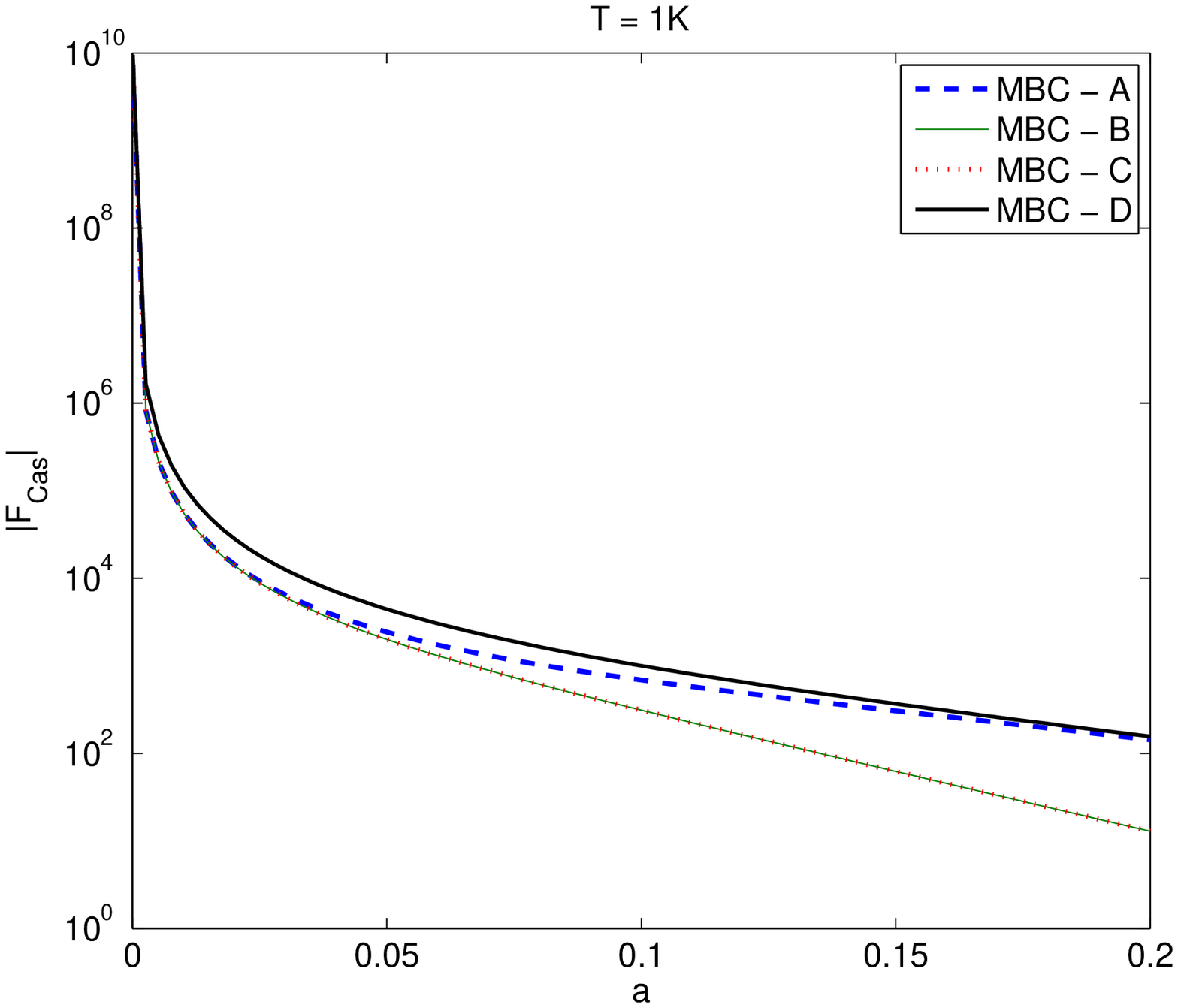}\epsfxsize=0.33\linewidth
\epsffile{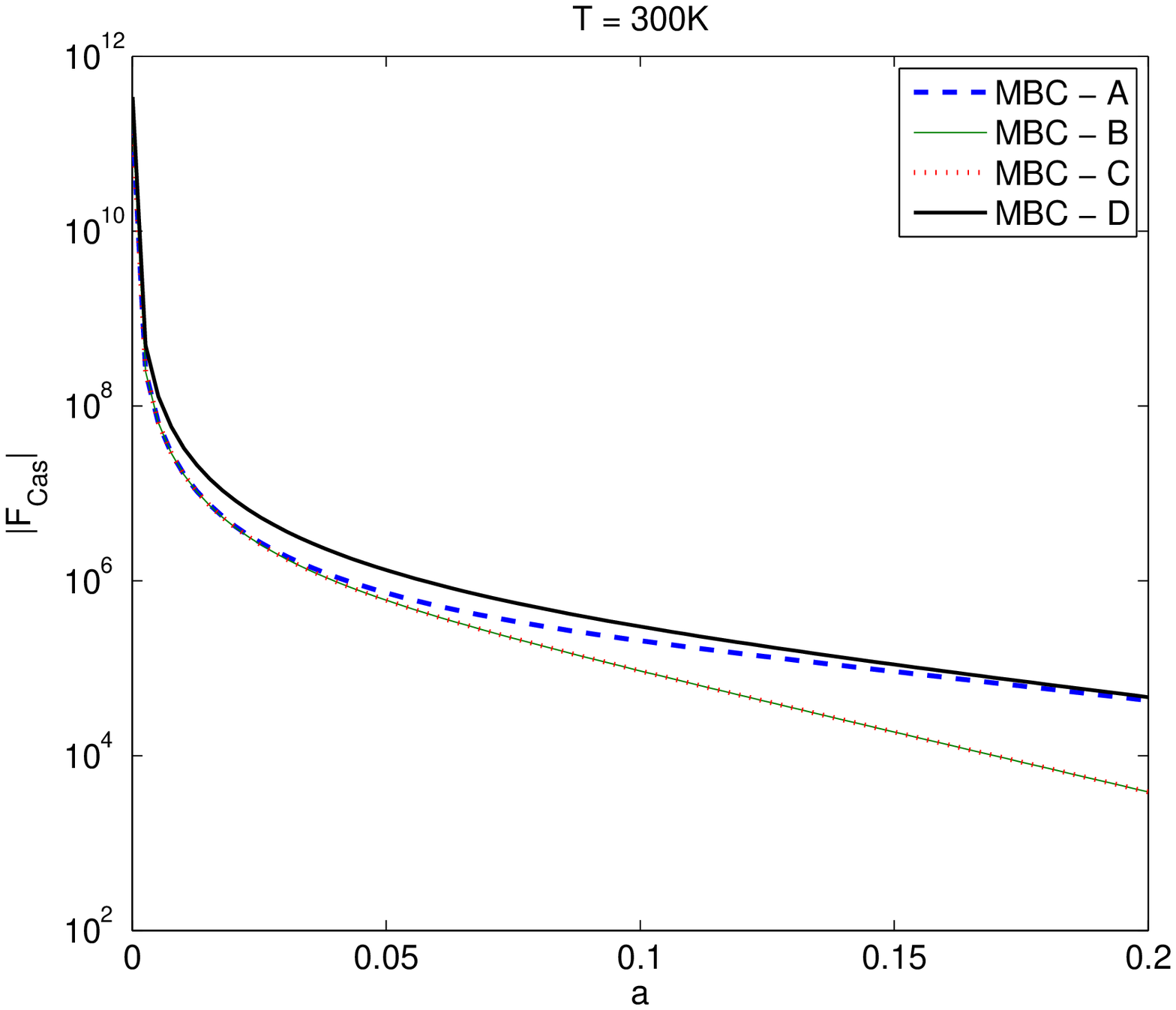}\caption{\label{Fig2} The absolute values of the   Casimir forces for various boundary conditions when $L_1\rightarrow \infty$ and $L_2=0.2\text{m}$. Here the unit of $a$ is m, the unit of  force is $\hbar c \times \text{N}$. The graphs correspond to $T=0\text{K}, 1\text{K}$ and $T=300\text{K}$ respectively. }\end{figure}

\begin{figure} \epsfxsize=0.5\linewidth
\epsffile{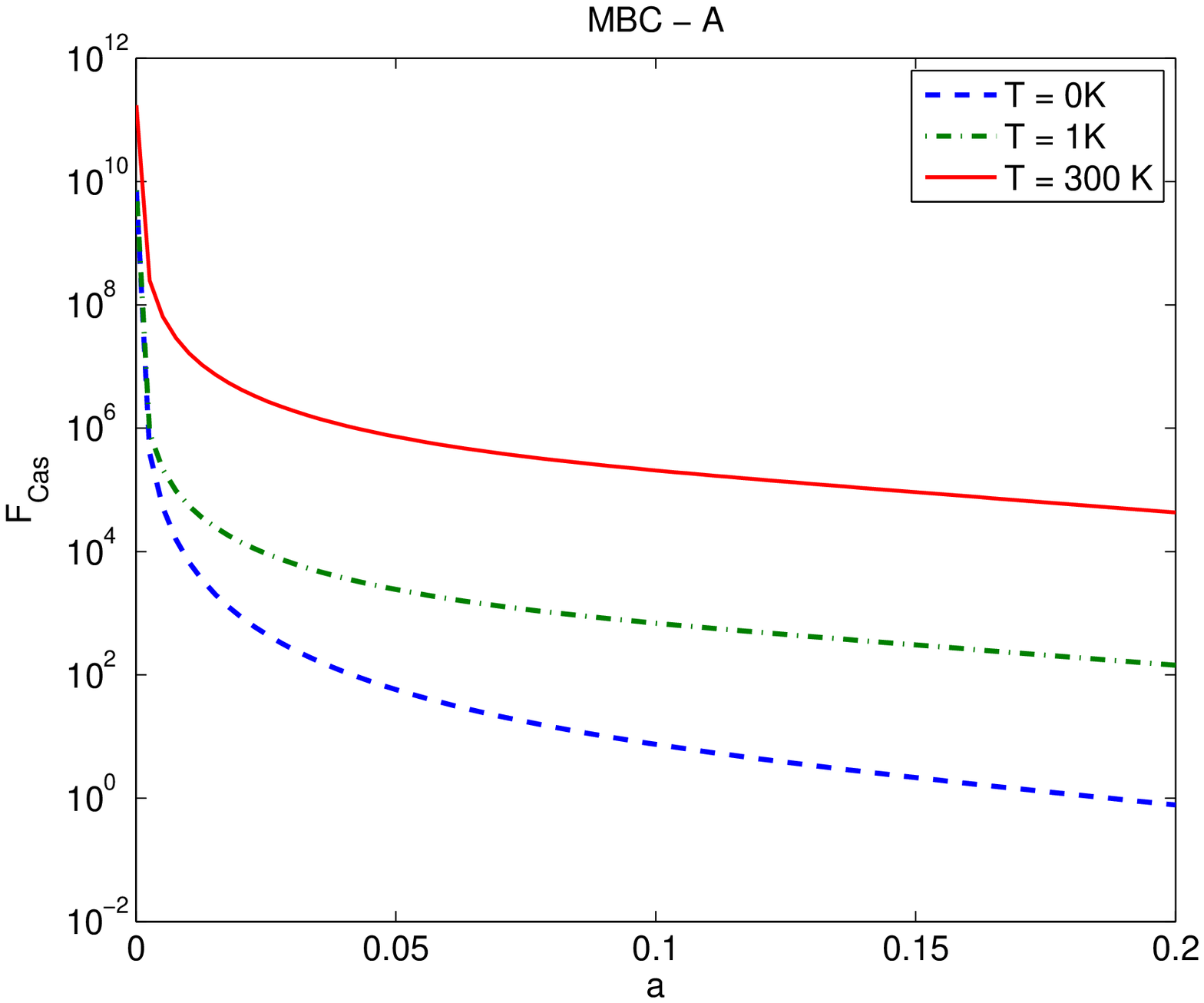}\epsfxsize=0.5\linewidth
\epsffile{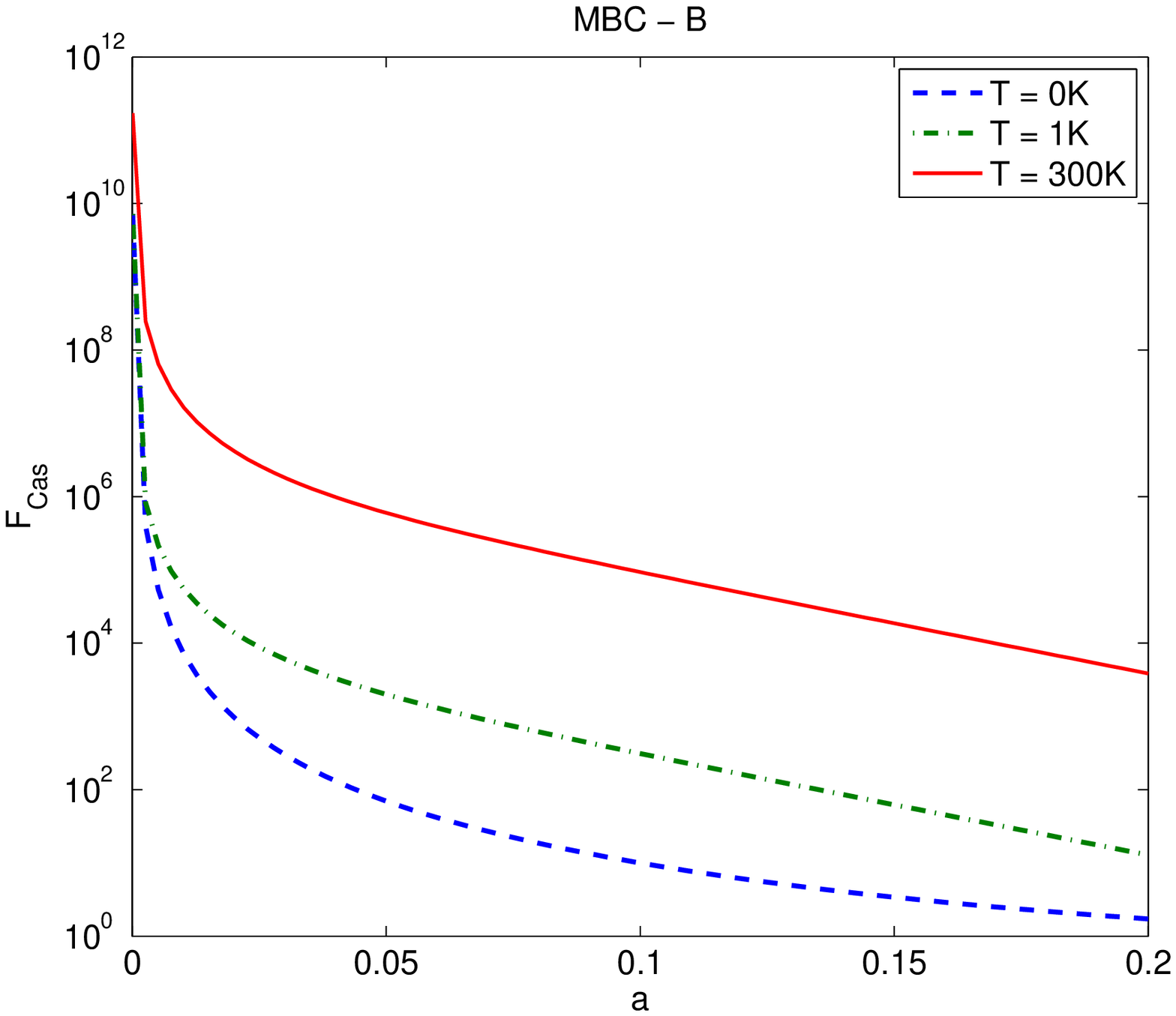}\\\epsfxsize=0.5\linewidth
\epsffile{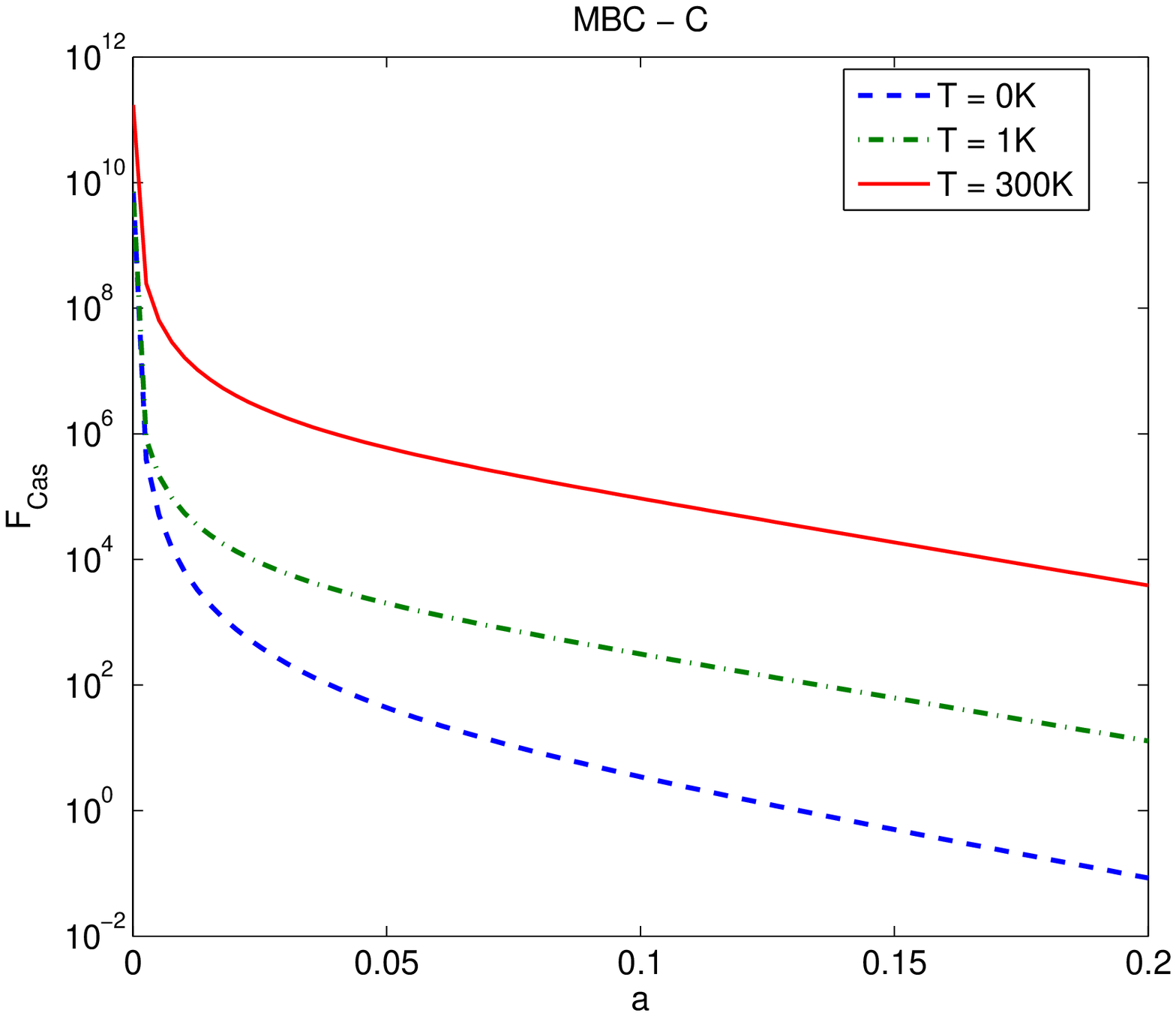}\epsfxsize=0.5\linewidth
\epsffile{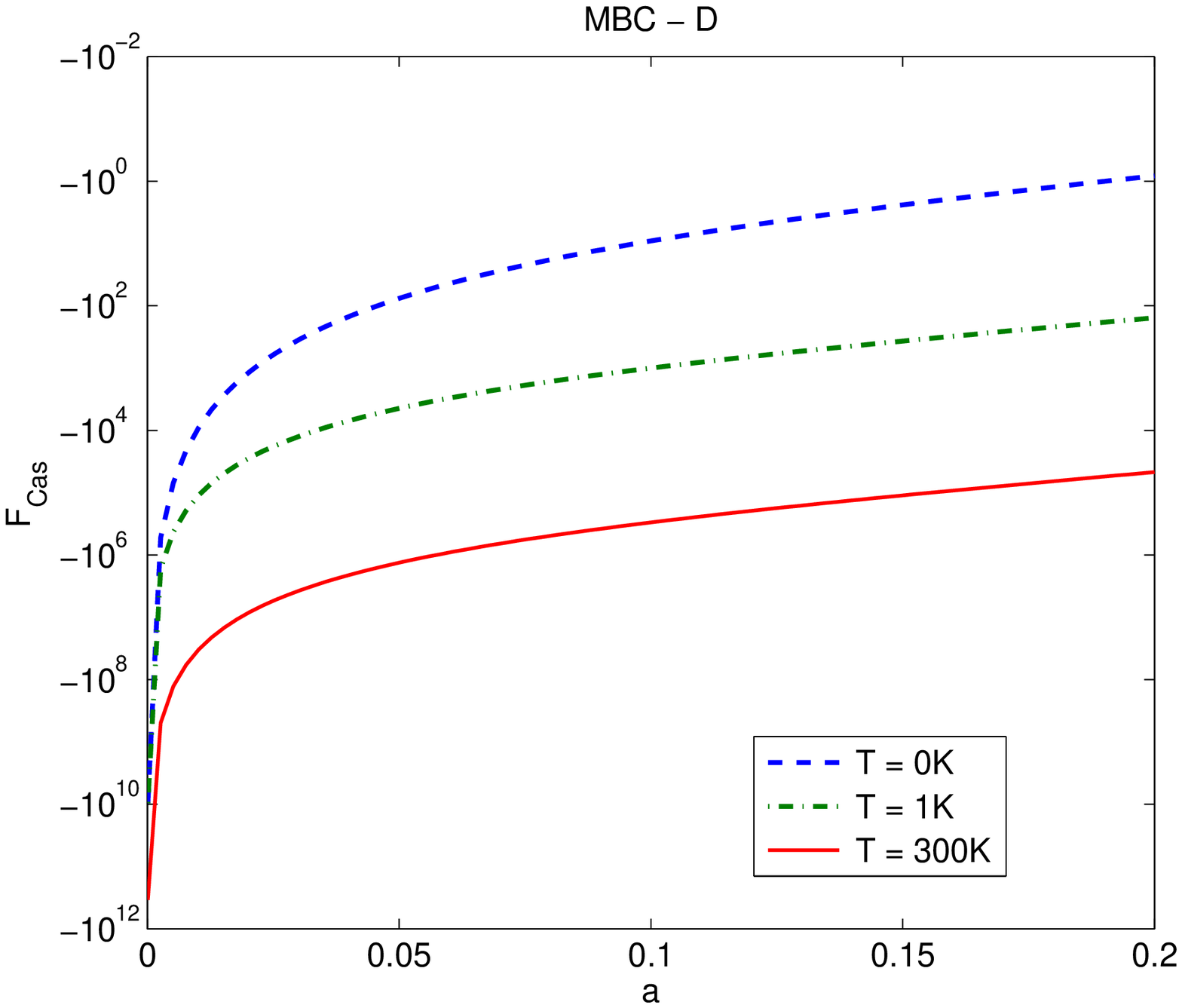}\caption{\label{Fig3} The  comparisons of the   Casimir forces at different temperatures when $L_1\rightarrow \infty$ and $L_2=0.2\text{m}$. Here the unit of $a$ is m, the unit of  force is $\hbar c \times \text{N}$. }\end{figure}

Comparing the magnitude of the Casimir force for various boundary conditions, we notice that in the case of open piston ($L_1\rightarrow \infty$), the magnitude of the Casimir force always decreases as the plate separation $a$ increases. Moreover, we see that when the plate separation $a$ is small, the leading term of the Casimir force for cases MBC-A, MBC-B and MBC-C which all assume mixed boundary conditions in the $x_1$ direction are the same and is equal to \begin{equation}\label{eq9_3_4}\frac{3\zeta_R(3)}{32\pi}\frac{L_2}{a^3}.\end{equation}For the cases MBC-D or MBC-E which assume pure boundary conditions in the $x_1$ direction, the leading term is \begin{equation}\label{eq9_3_5}-\frac{\zeta_R(3)}{8\pi}\frac{L_2}{a^3}.\end{equation}Its magnitude is $4/3$ times larger than the case of mixed boundary conditions in $x_1$ direction. \eqref{eq9_3_4} and \eqref{eq9_3_5} are also the corresponding zero temperature Casimir force in the infinite parallel plate limit.   On the other hand, we also notice that the classical limit of the Casimir force for the MBC-B and MBC-C cases are the same. For all the boundary conditions considered, the magnitude of the classical limit always decreases as the piston moves towards the equilibrium position. In the infinite parallel plate limit, the magnitude of the classical term for plates with mixed boundary conditions is half that for plates with pure boundary conditions. The comparisons of the Casimir forces with different boundary conditions and at different temperatures are depicted  in FIG. \ref{Fig2} and FIG. \ref{Fig3} respectively. Here we would like to remark that by restoring the units $\hbar$, $k_B$ and $c$, we have to replace $T$ in the expressions for Casimir force with $k_BT/(\hbar c)$. Therefore, for the physical $T=1\text{K}$ we need to substitute $T=436.7 \text{m}^{-1}$ which is actually   large if compared to $a$ in the range $0.01\text{m}\sim 0.2\text{m}$ which is equivalent to $a^{-1}$ in the range $5\text{m}^{-1}\sim 100 \text{m}^{-1}$. This explains the big difference between the zero temperature Casimir force and the Casimir force at $T=1\text{K}$ observed in FIG \ref{Fig3} when $a$ is in the range $0.01\text{m}\sim 0.2\text{m}$.  In fact, if we plot the Casimir force for $a$ in the range $< 0.1$mm, we would not observe  significant difference between the Casimir force at $T=0$K and $T=1$K.

This work can be generalized to higher dimensions, where the formulas is expected to be more complicated. Moreover, there will be more different combinations of boundary conditions. We leave this discussions to the future. Another interesting subject to explore is to consider a 'continuous' change   of boundary conditions from PEC b.c.~to PMC~b.c. on the piston but fixing the boundary condition on the opposite wall, and investigate the gradual change of the nature of the Casimir force on the piston. This may give us some insights into the mechanism of the change of the nature of Casimir force.

Finally, we would like to remark that although the piston scenario has the advantage of providing a formalism to obtain a Casimir force that is free of divergence problem, its has its own limitations. At the moment, this formalism cannot be used to obtain the Casimir force acting on the rectangular walls without substantial modification, otherwise it will lead to thermodynamically inconsistent Casimir effect. Some recent endeavors to  solve the problem of obtaining physically consistent Casimir force acting on the walls of a rectangular cavity can be found in \cite{nn1, nn2}. In particular, Geyer et al \cite{nn1} have proposed a formalism that can give a thermodynamically consistent Casimir energy in an ideal rectangular   metallic box.

\appendix
\section{Chowla--Selberg formula for  Epstein Zeta function and its derivative at zero}\label{appendix}
Here we gather some formulas we need for the Epstein zeta function \eqref{eq8_26_1} and its derivative at zero. The Chowla--Selberg formula \cite{6,7,8, 28, 29, 1_22_1, 1_22_2, 1_22_3, 1_22_4} says that
\begin{equation}\label{eq8_27_1}\begin{split}
&Z_{n}(s;c_1,\ldots, c_n)=Z_{m}(s; c_1, \ldots,
c_m)  +\frac{\pi^{m/2}\Gamma\left(s-\frac{m}{2}\right)}{\left[\prod_{j=1}^mc_j\right]\Gamma(s)}Z_{
n-m}\left(s-\frac{m}{2}; c_{m+1}, \ldots,
c_n\right)\\&+\frac{1}{\Gamma(s)}\frac{2\pi^{s}}{\left[\prod_{j=1}^mc_j\right]}\sum_{\boldsymbol{k}\in\widehat{\mathbb{Z}^m}\times \widehat{\mathbb{Z}^{n-m}}}\left(\frac{\sum_{j=1}^m\left[\frac{k_j}{c_j}\right]^2}
{\sum_{j=m+1}^n[c_{j}k_{j}]^2}\right)^{\frac{2s-m}{4}} K_{s-\frac{m}{2}}
\left(2\pi\sqrt{\left(\sum_{j=1}^{m}\left[
\frac{k_j}{c_j}\right]^2\right)\left(\sum_{j=m+1}^n[c_{j}k_{j}]^2\right)}\right), \end{split}
\end{equation} where $K_{\nu}(z)$ is the modified Bessel function.  By taking derivative with respect to $s$ and setting $s=0$, we find that
\begin{widetext}
\begin{equation}\label{eq8_27_3}\begin{split}
&Z_{n}'(0;c_1,\ldots, c_n) =Z_{m}'(0; c_1, \ldots,
c_m) +\frac{\pi^{-n/2}\Gamma\left(\frac{n}{2}\right)}{\left[\prod_{j=1}^nc_j\right]}  Z_{
n-m}\left(\frac{n}{2}; \frac{1}{c_{m+1}}, \ldots,
\frac{1}{c_n}\right)\\&+\frac{2}{\left[\prod_{j=1}^mc_j\right]}\sum_{\boldsymbol{k}\in\widehat{\Z^m}\times\widehat{\Z^{n-m}}} \left(\frac{\sum_{j=1}^m\left[\frac{k_j}{c_j}\right]^2}
{\sum_{j=m+1}^n[c_{j}k_{j}]^2}\right)^{-\frac{m}{4}}  K_{\frac{m}{2}}\left(2\pi\sqrt{\left(
\sum_{j=1}^m\left[\frac{k_j}{c_j}\right]^2
\right)\left(\sum_{j=m+1}^n[c_{j}k_{j}]^2\right)}\right).\end{split}
\end{equation}

\begin{acknowledgments}
This project is   supported by Ministry of Science, Technology and Innovation, Malaysia under e-Science fund 06-02-01-SF0080.
\end{acknowledgments}
\end{widetext}


\begin{thebibliography}{10}
\bibitem{1}
R. M. Cavalcanti, \emph{Casimir force on a piston}, Phys. Rev. D
\textbf{69} (2004), 065015.

\bibitem{13}
M. P. Hertzberg, R. L.  Jaffe, M. Kardar, A. Scardicchio,
\emph{Attractive Casimir forces in a closed geometry}, Phys. Rev.
Lett. \textbf{95} (2005), 250402.


\bibitem{14}
M. P. Hertzberg, R. L.  Jaffe, M. Kardar, A. Scardicchio,
\emph{Casimir forces in a piston geometry at zero and finite
temperatures}, Phys. Rev. D \textbf{76} (2007), 045016.

\bibitem{15}
V. N. Marachevsky, \emph{One loop boundary effects: techniques and
applications}, preprint arXiv: hep-th/0512221 (2005).

\bibitem{16}
G. Barton, \emph{Casimir piston and cylinder, perturbatively}, Phys.
Rev. D \textbf{73} (2006), 065018.

\bibitem{17}
V. N. Marachevsky, \emph{Casimir energy of two plates inside a
cylinder}, Phys. Rev. D \textbf{75} (2007), 085019.

\bibitem{18}
A. Edery, \emph{Casimir piston for massless scalar fields in three
dimensions}, Phys. Rev. D \textbf{75} (2007), 105012.

\bibitem{19}
A. Edery and I. Macdonald, \emph{Cancellation of nonrenormalizable
hypersurface divergences and the d-dimensional Casimir piston}, J.
High Energy Phys. \textbf{9} (2007), 0709:005.

\bibitem{20}X. H. Zhai and X. Z.Li, \emph{
Casimir pistons with hybrid boundary conditions}, Phys. Rev. D
\textbf{76} (2007), 047704.


\bibitem{21}
S. A. Fulling, L. Kaplan, and J. H. Wilson, \emph{Vacuum energy and
repulsive Casimir forces in quantum star graphs}, Phys. Rev. A
\textbf{76} (2007), 012118.

\bibitem{22}
V. N. Marachevsky, \emph{Casimir interaction: pistons and cavity},
J. Phys. A: Math. and Theor. \textbf{41} (2008), 164007.





\bibitem{23} A. Edery, V. N. Marachevsky, \emph{The perfect magnetic
conductor (PMC) Casimir piston in d+1 dimensions}, Phys. Rev. D \textbf{78} (2008), 025021.

\bibitem{24}
H. Cheng, \emph{The Casimir force on a piston in the spacetime with
extra compactified dimensions}, Phys. Lett. B \textbf{668} (2008), 72.



\bibitem{25}
S. C. Lim and L. P. Teo, \emph{Three dimensional Casimir piston for massive scalar fields}, preprint arXiv: hep-th: 0807.3613.




\bibitem{26}
S. C. Lim and L. P. Teo, \emph{Casimir piston at zero and finite temperature}, preprint arXiv: hep-th: 0808.0047, to appear in Eur. Phys. J. C.

\bibitem{27}
X. H. Zhai, Y. Y. Zhang and X. Z. Li, \emph{Casimir Pistons for Massive Scalar Fields},  preprint arXiv: hep-th:0808.0062.

\bibitem{5}
Steven~K. Blau, Matt Visser, and Andreas Wipf, \emph{Zeta functions
and the
  {C}asimir energy}, Nuclear Phys. B \textbf{310} (1988),  163.




\bibitem{6}
E.~Elizalde, S.~D. Odintsov, A.~Romeo, A.~A. Bytsenko, and
S.~Zerbini,
  \emph{Zeta regularization techniques with applications}, World Scientific
  Publishing Co. Inc., River Edge, NJ, 1994.

\bibitem{7}
Emilio Elizalde, \emph{Ten physical applications of spectral zeta
functions},
  Lecture Notes in Physics. New Series m: Monographs, vol.~35, Springer-Verlag,
  Berlin, 1995.


\bibitem{8}
K.~Kirsten, \emph{Spectral functions in mathematics and physics},
Chapman \&
  Hall/ CRC, Boca Raton, FL, 2002.

\bibitem{9}
E. Elizalde and A. Romeo, \emph{Expressions for the zeta--function
regularized {C}asimir energy}, J. Math. Phys. \text{30} (1989),
  1133.

\bibitem{10}
K. Kirsten, \emph{Casimir effect at finite temperature}, J. Phys. A
\textbf{ 24} (1991),  3281.

\bibitem{11}
G.~Ortenzi and M.~Spreafico, \emph{Zeta function regularization for
a scalar
  field in a compact domain}, J. Phys. A \textbf{37} (2004),
  11499.

 \bibitem{12}
S.C. Lim and L.P. Teo, \emph{Finite temperature Casimir energy in
closed rectangular cavities: a rigorous derivation based on zeta
function technique}, J. Phys. A: Math. Theor. \textbf{40} (2007),
11645.

\bibitem{1_15_1}J. Feinberg, A. Mann and M. Revzen,  \emph{Casimir effect: The classical limit},  Ann. Phys. \textbf{288} (2001), 103.

\bibitem{1_15_2} I. Klich, J. Feinberg, A. Mann A and M. Revzen, \emph{Casimir energy of a dilute dielectric ball with uniform velocity of light at finite temperature}, Phys. Rev. D \textbf{62} (2000), 045017.

\bibitem{1_15_3} M. Schaden and L. Spruch,    \emph{Classical Casimir effect: The interaction of ideal parallel walls at a finite temperature}, Phys. Rev. A \textbf{65} (2002), 034101.


\bibitem{1_15_4}A. Scardicchio and R.L. Jaffe,  \emph{Casimir effects: An optical approach II. Local observables and thermal corrections},  Nucl. Phys. B \textbf{743} (2006), 249.



  \bibitem{30}
Jan Ambj{\o}rn and S.~Wolfram, \emph{Properties of the vacuum. {I}.
  {M}echanical and thermodynamic}, Ann. Physics \textbf{147} (1983), 1.


 \bibitem{nn1} B. Geyer, G. L. Klimchitskaya and  V. M. Mostepanenko, \emph{Thermal Casimir effect in ideal metal rectangular boxes},   Euro. Phys. J. C. \textbf{57} (2008),  823.

     \bibitem{nn2} S. A. Fulling, L. Kaplan, K. Kirsten, Z. H. Liu and K. A. Milton, \emph{ Vacuum Stress and Closed Paths in Rectangles, Pistons, and Pistols}, preprint arXiv:0806.2468.


\bibitem{28}
S.~Chowla and A.~Selberg, \emph{On {E}pstein's zeta function. {I}},
Proc. Nat.
  Acad. Sci. U. S. A. \textbf{35} (1949), 371.

\bibitem{29}
A. Selberg and S.~Chowla, \emph{On {E}pstein's zeta-function}, J.
Reine
  Angew. Math. \textbf{227} (1967), 86.
\bibitem{1_22_1}M. Bordag, E. Elizalde and K. Kirsten, \emph{Heat kernel coefficients of the Laplace operator on the $D$-dimensional ball},  J. Math. Phys. \textbf{37} (1996), 895.
\bibitem{1_22_2}E. Elizalde, A. Romeo, \emph{Rigorous extension of the proof of zeta-function regularization}, Phys. Rev. D \textbf{40} (1989), 436.
\bibitem{1_22_3}
E. Elizalde, \emph{Multidimensional extension of the generalized Chowla-Selberg formula}, Commun. Math. Phys. \textbf{198} (1998), 83.
\bibitem{1_22_4}
E. Elizalde, \emph{Explicit zeta functions for bosonic and fermionic fields on a non-commutative toroidal spacetime},
 J. Phys. A \textbf{34} (2001), 3025.

\end{thebibliography}
\end{document}